\documentclass[lettersize,journal]{IEEEtran}
\usepackage{amsmath,amsfonts}
\usepackage{algorithmic}
\usepackage{algorithm}
\usepackage{array}
\usepackage[caption=false,font=normalsize,labelfont=sf,textfont=sf]{subfig}
\usepackage{textcomp}
\usepackage{stfloats}
\usepackage{url}
\usepackage{verbatim}
\usepackage{graphicx}
\usepackage{cite}
\usepackage{physics}
\usepackage {tabularx} 
\usepackage{amssymb}
\usepackage{amsmath}
\usepackage[]{units}
\usepackage[]{upgreek}
\DeclareUnicodeCharacter{0308}{\"}
\DeclareUnicodeCharacter{030C}{\v{ }}
\DeclareUnicodeCharacter{0301}{\'{ }}

\begin{document}
\title{Reversal of coupled vortices in advanced spintronics: A mechanistic study}

\author{Abbass Hamadeh, Abbas Koujok, Salvatore Perna, Davi R. Rodrigues, Alejandro Riveros, Vitaliy Lomakin, Giovanni Finocchio, Grégoire de Loubens, Olivier Klein and Philipp Pirro
\thanks{A. Hamadeh (corresponding author),  A. Koujok, and  P. Pirro  are with Fachbereich Physik and Landesforschungszentrum OPTIMAS, Rheinland-Pf\"alzische Technische Universit\"at Kaiserslautern-Landau, 67663 Kaiserslautern, Germany.(e-mail: hamadeh@rhrk.uni-kl.de)}
\thanks{S. Perna is with Department of Electrical Engineering and ICT, University of Naples Federico II, Naples, Italy}
\thanks{D. R. Rodrigues is with the Department of Electrical and Information Engineering,
Politecnico di Bari, Bari, 70126 Italy.}
\thanks{A.Riveros is with Escuela de Ingeniería, Universidad Central de Chile, 8330601, Santiago, Chile}
\thanks{V. Lomakin is with Center for Magnetic Recording Research, University of California San Diego, La Jolla, California 92093-0401, USA.}
\thanks{G. Finocchio is  with the Department of Mathematical and Computer Sciences, Physical Sciences and Earth Sciences, University of Messina, I -98166, Messina, Italy.}
\thanks{G. de Loubens is with SPEC, CEA, CNRS, Universit\'e Paris-Saclay, 91191 Gif-sur-Yvette, France.}
\thanks{O. Klein is with Univ. Grenoble Alpes, CEA, CNRS, Grenoble INP, Spintec, 38054 Grenoble, France.}
}

\maketitle

\begin{abstract}
This study conducts a comprehensive investigation into the reversal mechanism of magnetic vortex cores in a nanopillar system composed of two coupled ferromagnetic dots under zero magnetic field conditions. The research employs a combination of experimental and simulation methods to gain a deeper understanding of the dynamics of magnetic vortex cores. The findings reveal that by applying a constant direct current, the orientation of the vortex cores can be manipulated, resulting in a switch in one of the dots at a specific current value. The micromagnetic simulations provide evidence that this switch is a consequence of a deformation in the vortex profile caused by the increasing velocity of the vortex cores resulting from the constant amplitude of the trajectory as frequency increases. These findings offer valuable new insights into the coupled dynamics of magnetic vortex cores and demonstrate the feasibility of manipulating their orientation using direct currents under zero magnetic field conditions. The results of this study have potential implications for the development of vortex-based non-volatile memory technologies.
\end{abstract} 


\section{\textbf{Introduction}}
\IEEEPARstart{I}{n }a magnetic medium, a vortex state describes a region of in-plane curling magnetization distribution spiraling about a region with out of plane magnetization, namely the vortex core (VC). Magnetization patterns at the vortex state can be described by the vortex polarity and chirality. The vortex polarity describes the polarization of the magnetization at the VC, where a positive/negative polarity refers to an upward/downward direction of magnetization respectively. On the other hand, the chirality describes the direction of the curling magnetization about the VC, with a positive/negative chirality referring to a clockwise/counter clockwise configuration. In the past years, a number of theoretical, numerical and experimental studies have been devoted to understanding and manipulating the vortex dynamics \cite{guslienko2002eigenfrequencies,choe2004vortex,zhu2005broadband,novosad2005magnetic,guslienko2006magnetic,lee2007gyrotropic,pigeau2011optimal,dussaux2012field,guslienko2014nonlinear,ramasubramanian2020tunable}. Magnetic vortices offer high density data storage application possibilities, such as non-volatile magnetic memories \cite{sousa2005non,vsoltys2013magnetic}. These solitons also possess the ability to be exploited further as sources for spin wave generation \cite{wintz2016magnetic,mayr2021spin}, which constitutes the backbone for beyond-CMOS-based technologies \cite{khitun2010magnonic,kruglyak2010magnonics,chumak2015magnon,hamadeh2022hybrid}.

In recent years, there has been increasing interest in studying vortex dynamics in the context of spin-transfer torque oscillators (STOs) \cite{dussaux2012field,guslienko2014nonlinear,ramasubramanian2020tunable,hamadeh2022role}. These non-linear devices have the ability to regulate their frequencies as to synchronize with external stimuli, such as current or field \cite{bonin2009analytical,dussaux2011phase,hamadeh2014perfect}, noise in the surrounding environment \cite{nakada2012noise}, or other STOs \cite{locatelli2015efficient,lebrun2017mutual}. This nominates them as promising candidates for applications in vowel recognition \cite{romera2018vowel}, neural networks \cite{riou2019temporal}, and neuromorphic computing \cite{torrejon2017neuromorphic}. Vortices can be excited and driven in the magnetic layers of STOs by applying a DC current (I$_{\mathrm{DC}}$) through the multi-layered structure, a phenomenon known as spin-transfer torque (STT) \cite{slonczewski1996current,li2003magnetization,shibata2006current}. Compared to traditional STOs, vortex-based STOs offer improved efficiency \cite{belanovsky2012phase,locatelli2015efficient}, a wider range of tunability \cite{hamadeh2014origin}, and a narrower frequency linewidth \cite{ruotolo2009phase,pribiag2007magnetic,locatelli2011dynamics}. One of the most significant applications of vortex-based STOs is their ability to synchronize more efficiently and achieve mutual phase locking \cite{dussaux2011phase,belanovsky2012phase,hamadeh2014perfect,araujo2015optimizing,locatelli2015efficient,lebrun2017mutual,li2018selective,zeng2021synchronization,hamadeh2022role}.

In this work, we investigate the dynamics of two dipolarly coupled vortices for continuous applied DC current and zero field, while also aiming to reveal the origin of the reversal mechanism underwent by the VC based in the STO's thick layer. Understanding the physics bound to the origin of the VC's reversal mechanism in coupled systems is at the core of studies aiming at vortex engineering, control and manipulation. The concept of VC reversal itself is neither new nor surprising, as it has been previously approached and investigated via various methods \cite{van2006magnetic,yamada2007electrical,kim2007electric,locatelli2013reversal}. However, interpretations in regards of the VC's dynamic reversal in coupled systems under zero magnetic field  still lag as to vividly convey the nature of such mechanism. Here, we propose an alternative approach as to vortex coupled dynamics for applied DC current at zero field, thus presenting an efficiently convenient approach to controlling vortex dynamics in coupled ferromagnetic dots. After which, we go further to provide  detailed understanding of the dynamic VC reversal mechanism in such coupled systems.

\section{\textbf{Experimental Setup}}

\begin{figure}[!h]
\centerline{\includegraphics[width=0.5\textwidth ]{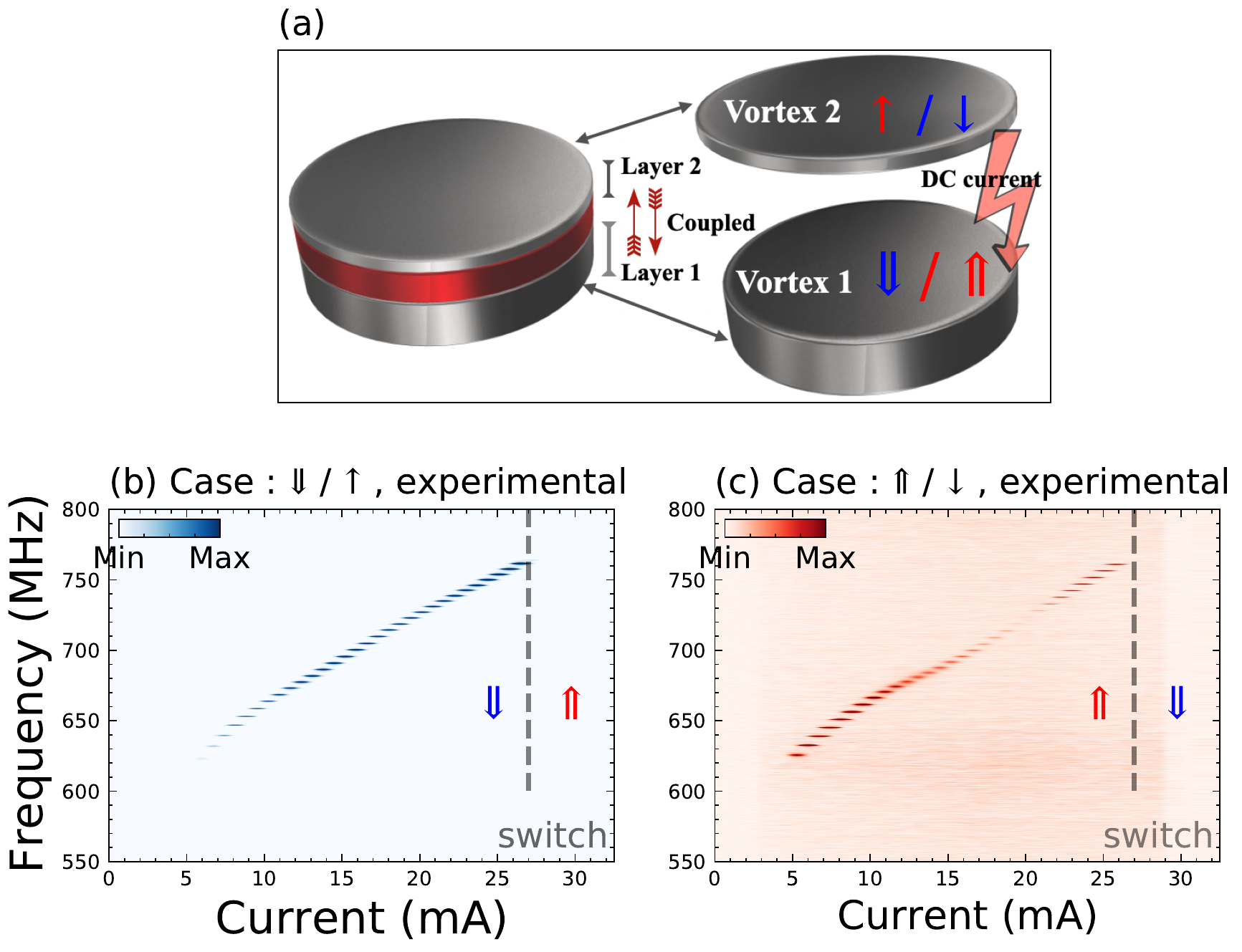}}
\caption{Schematics of the coupled vortex STO device. (a) illustrates the device structure, consisting of two Permalloy layers (Layer 1 and 2) of thicknesses 15 nm and 4 nm, respectively, separated by a non-magnetic copper spacer of thickness \unit[10]{nm}. The diameter of each disc is \unit[250]{nm}. The injected DC current (indicated by the large red arrow) flows downwards through the structure. The vortices in both layers have opposite polarities and same chiralities. (b) and (c) show the dependence of the microwave generation characteristics of the STO on the applied DC current at a zero bias field, for two different configurations of vortex polarities: (b) Case 1: $\Downarrow$/$\uparrow$ and (c) Case 2: $\Uparrow$/$\downarrow$.}
\label{fig1}
\end{figure}

To start, we present the geometry of the circular spin-valve nanopillar oscillator under investigation (see Figure \ref{fig1}(a)). The system is a multi-layered stack, consisting of Permalloy-based thick (vortex 1) and thin (vortex 2) layers with vortex magnetic configurations at remanence. The vortices' configurations in these layers are prepared with opposite polarities and the same chiralities (as shown in Figure \ref{fig1}(a) for case 1 and case 2), where the polarities of the vortices in the thick layer are represented by $\Uparrow$ or $\Downarrow$, and those in the thin layer by $\uparrow$ or $\downarrow$. The thickness of the lower layer is \unit[15]{nm}, while that of the upper layer is \unit[4]{nm}. A non-magnetic copper spacer of \unit[10]{nm} thickness is present between the two layers. The overall diameter of the multi-layered stack is \unit[250]{nm}.

We initiate our study by experimentally investigating the power dependence of the thick layer's frequency on DC current in the absence of applied magnetic field for the two above mentioned cases (see Fig. \ref{fig1}(b) and \ref{fig1}(c)). As I$_{\mathrm{DC}}$ from \unit[5]{mA} to \unit[27]{mA}, the fundamental frequency of the thick layer's VC starts increasing for both configurations $\Downarrow/\uparrow$ and $\Uparrow/\downarrow$ (see Fig. \ref{fig1}(b) and \ref{fig1}(c)). The increase of which is owed to increasing the Zeeman energy associated with the circumferential Oersted field generated by the injected current \cite{pribiag2007magnetic}. As such, the vortex exhibiting stronger confinement increases its frequency from \unit[600]{MHz} until reaching a maximum value of \unit[770]{MHz} for an applied current of \unit[27]{mA}. Increasing the applied current I$_{\mathrm{DC}}$ further, dynamics disappear and the signal's power becomes null. Experimentally, hindered dynamics refer to coupled vortices with same polarities. For opposite core polarities, the thick layer polarizes electrons during I$_{\mathrm{DC}}$ traversal, polarized spins thereafter transfer spin angular momentum onto magnetization in the thin layer resulting in the observed dynamics. However, this exerted torque by the polarized spins becomes absent as the thick layer's VC switches its orientation. Thus, the region with no dynamics (see dotted switching region, Fig. \ref{fig1}(b) and \ref{fig1}(c)) represents the thick layer's VC orientation reversal at which the signal's power is null. This intriguing behavior of the thick layer's VC's coupled dynamics arises a fundamental question about the origin of the dynamic reversal itself. The origin of which has neither been explained nor revealed for a system based on coupled vortex dynamics.

\section{\textbf{Micromagnetic Simulations: Results and Analysis}}

\begin{figure}[!h]
\centerline{\includegraphics[width=0.5\textwidth ]{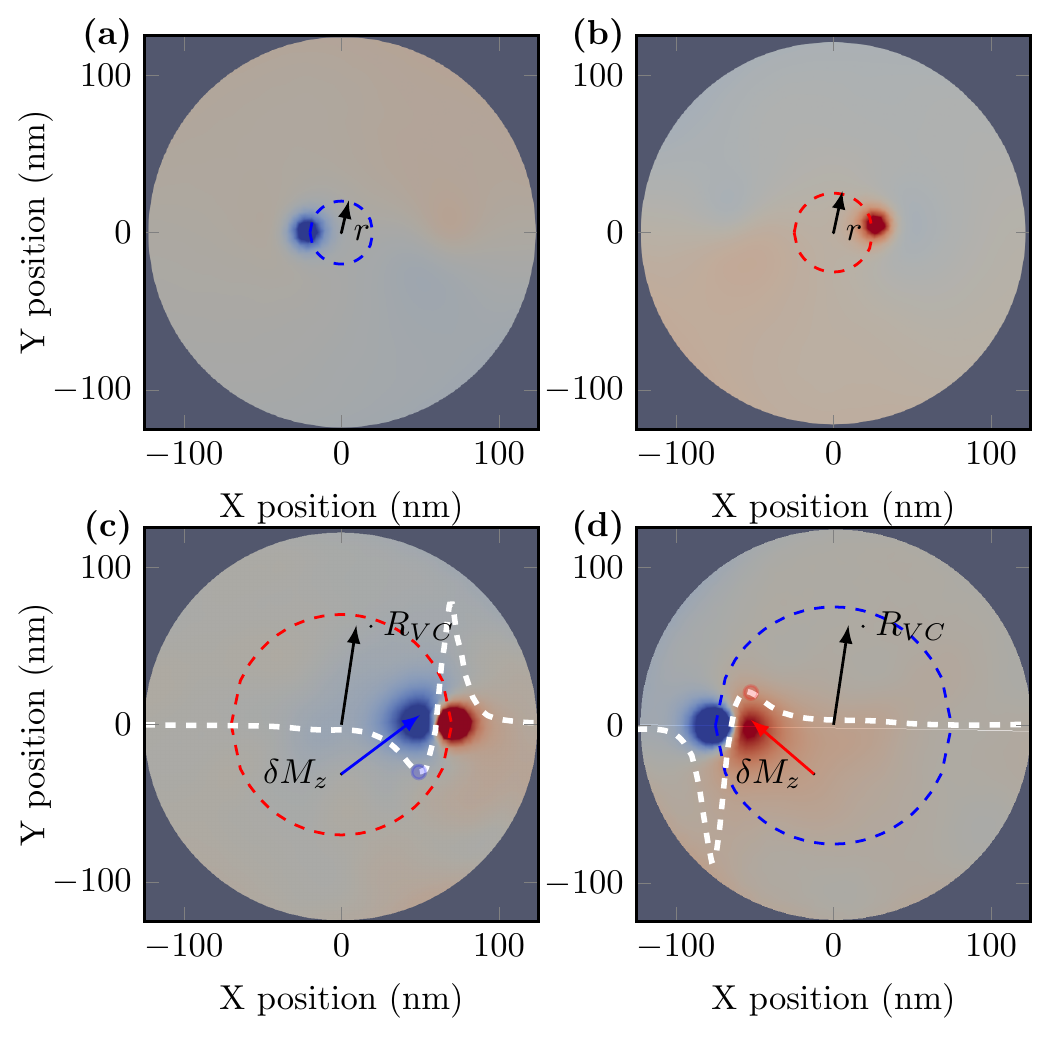}}
\caption{Sketch of the out of plane component of the vortex magnetization M$_z$ and the trajectories of the VCs of each layer as well as their respective radii. $\Uparrow$/$\downarrow$ configuration (a, c); $\Downarrow$/$\uparrow$ configuration (b, d). Thin layer's VC trajectory and its respective radius r (a, b). Thick layer's VC trajectory and its respective radius R$_\text{VC}$(c, d). Figures (c, d) also present the VC magnetization (dashed white line) profiles along the line crossing the VC and the nanodot, in addition to $\delta M_\text{z}$ whose direction opposes that of the VC magnetization.}
\label{fig2}
\end{figure}

In the following we go further to reveal this origin, as well as explain vividly the dynamic reversal mechanism observed in such devices. Via micromagnetic simulations, we investigate the coupled dynamics of the thin and thick layers' VCs for the vortex based oscillator stack introduced earlier in Fig. \ref{fig1}(a). To carry out the simulations and the necessary data analysis, we use FastMag \cite{chang2011fastmag} and the software platform Aithericon \cite{aithericon}. As such, we utilize the same oscillator dimensions presented earlier, in addition to structure parameters from the experimental work \cite{naletov2011identification}. We monitor the dynamics of the two coupled vortices tracing the trajectories of their respective VCs (see Fig. \ref{fig2}). Clearly, it is noticeable that the radius of the trajectory followed by the thin layer's VC is much smaller than that of the thick layer for both cases $\Downarrow/\uparrow$ (Fig. \ref{fig2}(b,d)) and $\Uparrow/\downarrow$ (Fig. \ref{fig2}(a,c)). Focusing on (Fig. \ref{fig2}(c,d)), a region on the inner side of the thick layer's VC is observed. The region of which has a normalized  amplitude $\delta M_\text{z}$, and exhibits a polarity opposing that of the thick layer's VC.

\begin{figure}[!h]
\centerline{\includegraphics[width=0.5\textwidth]{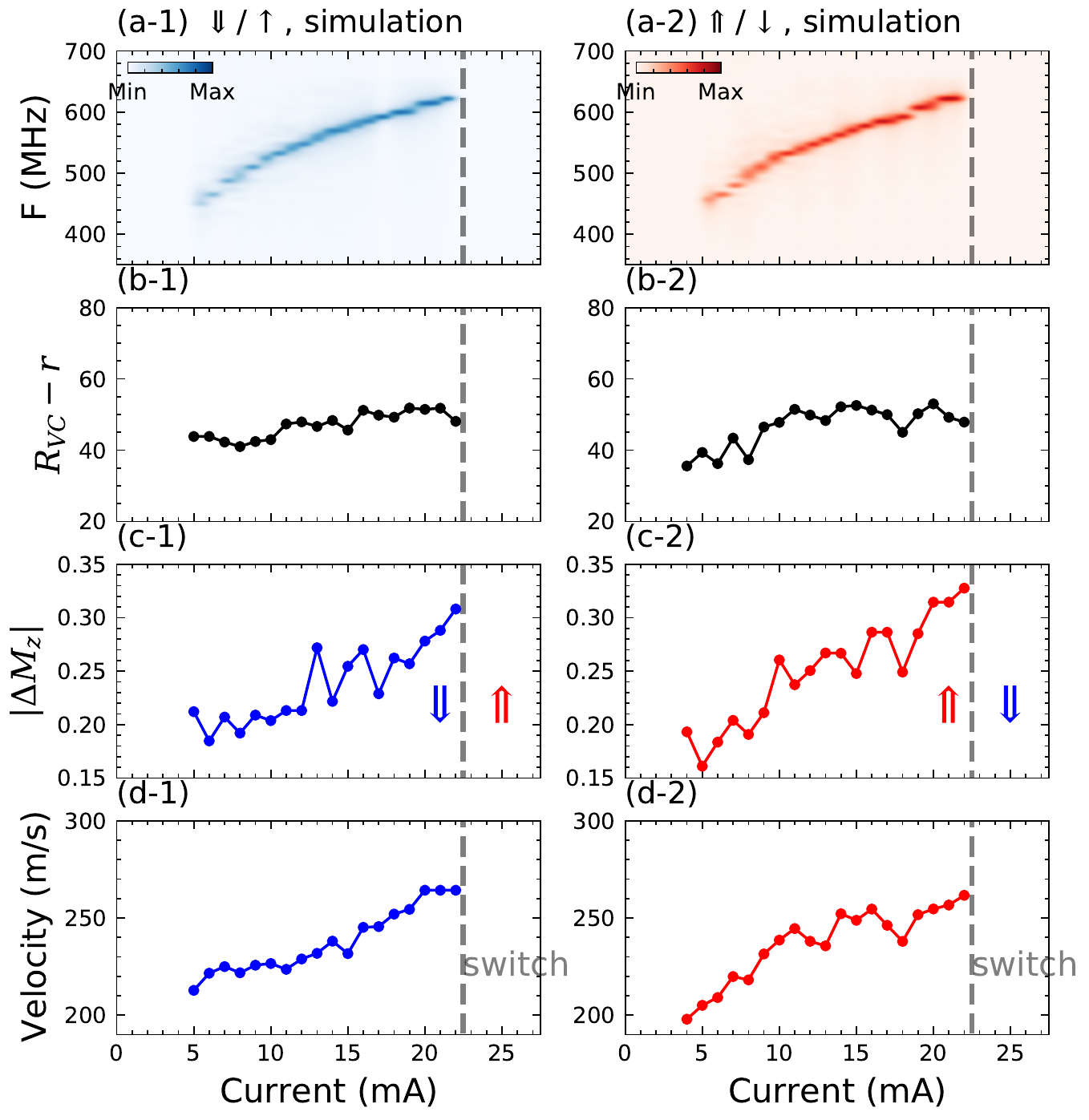}}
\caption{Power spectrum of the frequency (in MHz) extracted from the real part of the Fast Fourier Transformation on $M_\text{x}$+i$M_\text{y}$, the difference between the thick and thin layer's radii, the deformation amplitude, and the velocity (in m/s) of the moving thick layer's VC are plotted versus current (in mA) in the first, second, third, and fourth rows respectively, for (1) $\Downarrow$/$\uparrow$ polarities and (2) $\Uparrow$/$\downarrow$ polarities. VC orientation reversal occurs at the maximum velocity and deformation amplitude values (dotted switching region).}
\label{fig3}
\end{figure}

To explain the appearance of $\delta M_\text{z}$, we study the thick layer's VC frequency dependence on applied DC current in the absence of an external field (Fig. \ref{fig3}). The current traversing the structure downwards in both configurations ($\Downarrow/\uparrow$ and $\Uparrow/\downarrow$) excites a high-frequency orbital motion of the VC, and induces an Oersted field of clockwise direction. The Oersted field increases the tunable frequency range \cite{zaspel2019high}.

Upon increasing the current gradually above \unit[5]{mA}, the system's fundamental frequency starts increasing (see power spectrum Fig. \ref{fig3}(a)) for both $\Downarrow/\uparrow$ and $\Uparrow/\downarrow$) \cite{khvalkovskiy2009vortex}, raising the velocity and $\delta M_\text{z}$ at the VC. Here, it should be noted that the increase in velocity of the thick layer's VC may be traced back to either an increase in the fundamental frequency, or a decrease in its trajectory's radius. However, the thick layer's radius is observed to vary only slightly with respect to that of the thin layer (see Fig. \ref{fig3}(b)). This goes back to the fact that the two VCs are coupled, and thus can't change their trajectories' radii freely, as the system consisting of the dipolarly coupled vortices exhibits a selection upon the radius to radius distance \cite{guslienko2005dynamics}. This implies that, on average R$_\text{VC}$-r lies in the vicinity of a somewhat constant value. Whence, the increase in the system's fundamental frequency is associated with an increase in the velocity of the thick layer's VC (see Fig. \ref{fig3}(d)). Naturally, the VC being a topological soliton deforms upon a continuous increase in its velocity, this is made clear by the appearance
of the previously mentioned region on the inner side of the thick layer's VC (see Fig. \ref{fig2}(c,d) for $\delta M_\text{z}$). The respective dip and hump shaped regions represent the mentioned deformation at the inner side of the thick layer's VC during its gyrotropic motion about its equilibrium position for both cases, each of which has an amplitude of $\delta M_\text{z}$ (see Fig. \ref{fig2}(c) dip, and Fig. \ref{fig2}(d) hump). 

As the current reaches \unit[22.5]{mA}, the vortex attains a critical velocity of \unit[266]{m/s} (see Fig. \ref{fig3}(d)). Beyond this value, the VC reverses its orientation (see the switching region in Fig. \ref{fig3}(c) and Fig. \ref{fig3}(d) for both $\Downarrow/\uparrow$ and $\Uparrow/\downarrow$), leading to the anticipated velocity drop \cite{guslienko2008dynamic}. The results of the micromagnetic simulations are in qualitative agreement with our obtained experimental results. Quantitatively, the difference between the frequency measured experimentally and that measured via simulations is approximately $\abs{\Delta F}$ = \unit[100]{MHz}. The respective difference is rather expected based on the fact that thermal fluctuations haven't been accounted for in the simulations. The latter's influence on frequency in such systems has been previously studied and demonstrated \cite{d2017effect}. In addition to the mentioned thermal influence, the simulation's accounted for Oersted field has been calculated analytically for the case of a DC current traversing an infinitely long wire, then plugged as part of the effective field. 

Hereby, the value of \unit[266]{m/s} reached constitutes a critical velocity of the VC, where any further increase in applied current implies a VC reversal. The obtained critical velocity is in good agreement with the analytical expression \cite{khvalkovskiy2010critical} V$_\text{c}$= (1.66±0.18)$\times \gamma \times \sqrt{A_\text{ex}}$ (e.g V$_\text{c}^{Py}$= \unit[289$\pm$31]{m/s}), where $\gamma$= $5.51\times10^{7}$ T$^{-1}s^{-1}$ is the gyromagnetic ratio, and A$_\text{ex}$= \unit[1$\times 10^{-11}$]{J/m} is the exchange stiffness. The deformation of amplitude $\delta M_\text{z}$ initially stated in Fig. \ref{fig2}(c) and Fig. \ref{fig2}(d) opposes the polarity of the undeformed vortex core. For a positive VC polarity ($\Uparrow$) this deformation appeared as a dip, whereas it appeared as a hump for a negative VC polarity ($\Downarrow$). $\abs{\Delta M_\text{z}}$ increases in a similar manner to the velocity/frequency increase with current between \unit[5]{mA} and \unit[22.5]{mA} (see Fig. \ref{fig3}(c) for $\abs{\Delta M_\text{z}}$ vs. current for both $\Downarrow/\uparrow$ and $\Uparrow/\downarrow$). This dip's/hump's amplitude increase reaches a maximum at the critical velocity of \unit[266]{m/s}, at which the deformation in the VC becomes large enough to induce an orientation reversal ($\Downarrow$ to $\Uparrow$ in \ref{fig3}(c-1) and $\Uparrow$ to $\Downarrow$ in \ref{fig3}(c-2)), leading to a drop in the VC's velocity accompanied by spin waves' emission \cite{choi2007strong}.

\section{\textbf{Conclusion}}

The results of our experimental studies and micromagnetic simulations, along with the physical interpretations provided, offer a comprehensive understanding of the relationship between the dynamics of coupled vortices and applied DC current. Our research has clarified the dynamic reversal mechanism, which is traced to the appearance of a dip/hump region on the inner side of the vortex core in the thick layer due to increasing velocity of the vortex core. The continuously increasing velocity of the vortex motion in response to the applied DC current is shown by the emergence of deformation. Our findings provide significant insights into the engineering of vortex switching in nano-oscillators, enabling the manipulation of magnetization dynamics and control of vortex polarity. The vortex trajectories of coupled layers present many possibilities for engineering amplitude and velocity of VCs.This study provides a significant contribution to a deeper understanding of the physics behind vortex core reversal in coupled systems without a static magnetic field, which is crucial for the advancement of advanced spintronic devices.

\section*{\textbf{Acknowledgments}}

This work has been  supported  by the European Research Council within the Starting Grant No. 101042439 "CoSpiN" and by the Deutsche Forschungsgemeinschaft (DFG, German Research Foundation) - TRR 173 - 268565370" (project B01).

\bibliographystyle{IEEEtran}
\bibliography{references}

\newpage

\vfill

\end{document}